\documentclass[12pt]{article}
\usepackage{amsmath}
\usepackage{graphicx}
\usepackage{amsfonts}
\usepackage{amssymb}
\usepackage{epsfig}

\newcommand{\ap}{{\alpha^{\prime}}}

\newcommand{\esp}{{\rm e}}

\newcommand{\no}{\nonumber}

\newcommand{\bnp}[1]{Nucl. Phys. {\bf #1}}
\begin{document}
\begin{titlepage}
\rightline{DFTT-22/2001}
\vfill \centerline{\Large\bf The enhan\c{c}on mechanism for fractional branes $^\ast$}
\vfill \centerline{\bf Paolo Merlatti} \centerline{\sl Dipartimento di Fisica Teorica,
Universit\`a di Torino} \centerline{\sl and I.N.F.N., sezione di Torino} \centerline{\sl
via P. Giuria 1, I-10125 Torino, Italy} \centerline{\sl Paolo.Merlatti@to.infn.it}\vfill
\begin{abstract}
We study the enhan\c{c}on mechanism for fractional D-branes in conifold and orbifold
backgrounds and show how it can resolve the repulson singularity of these geometries. In
particular we show that the consistency of the excision process requires that the interior
space be not empty. In the orbifold case, we exploit the boundary state formalism to
obtain an explicit conformal description and emphasize the non trivial role of the volume
of the internal manifold.
\end{abstract}

\vspace{2mm} \vfill \hrule width 3.cm {\footnotesize
 $^ \ast $ \hskip 0.1cm Research partially supported by the EC  RTN programme
HPRN-CT-2000-00131}
\end{titlepage}
\tableofcontents
\section{Introduction}

In the context of the gauge/gravity duality and to better understand the non-perturbative
dualities  between different string theories, a lot of effort has been recently devoted to
the study of fractional D-branes on conifold and orbifold singularities. Indeed the
introduction of fractional branes breaks conformal invariance and introduces RG flow
\cite{RG}, thus describing more ``realistic'' gauge theories
\cite{conifold,KT,KS,gauge,bertolini}.

It is well known that we lack of an explicit description, in terms of two-dimensional
conformal field theory, of ordinary and fractional D$p$-branes on a conifold background;
therefore many techniques such as the boundary state formalism or the brane-probe
computations, cannot be applied in this case. Obviously it is still possible to construct
classical supergravity solutions, describing space-filling branes in such a background.
However those solutions are singular \cite{herzog} at a certain ``small'' value of the
radial coordinate of the cone (in the IR of the dual gauge theory).

The appearance of naked singularities in the gravitational dual of non-conformal gauge
theories with reduced supersymmetry is a rather common phenomenon. In the past year we
learned that these naked singularities are potentially related to interesting IR dynamics
of their dual gauge theories. Thus, understanding the physics of these singularities is an
important problem because it could teach us about non-perturbative effects in the gauge
theory.

From a supergravity point of view, we can think that, at the typically short distances
where these singularities occur, some stringy mechanism comes into play, so that
supergravity solutions cease to make sense. However in this paper we show that, rather
surprisingly, pure supergravity tools let us to construct non-singular solutions also for
these conifold spaces, with the same techniques used in \cite{myers} for the case of a
$K3$ background. We'll find that it's possible to have a well defined solution if the
fractional branes, instead of being placed at the origin, form a composite shell,
extending along some directions of the base of the cone, at a radius $r\geq r_*$, where
$r_*$ is a small fixed radius we will determine in each case we will discuss. This is very
similar to what happens in the enhan\c{c}on mechanism \cite{pol,myers}. Indeed we think
that also in the case of the conifold background an enhan\c{c}on mechanism takes place. In
this case the base of the cone at $r=r_*$ defines the enhan\c{c}on locus and we argue that
the fractional branes become massless there.

We will show these results in the case of fractional D$p$-branes with $p\leq 3$. We will
particularly emphasize the $p=3$ case because of its relevance on the gauge theory side
\cite{conifold,KT,KS,buchel,non-extremal,restoration}. This case has already been
extensively studied from different perspectives: in \cite{KT}, Klebanov and Tseytlin found
a solution that exhibits a singularity in the IR. In a subsequent paper \cite{KS} Klebanov
and Strassler, getting insights from the strong dynamic of the infrared gauge theory,
argued that the naked singularity of the Klebanov-Tseytlin (KT) geometry should be
resolved via the deformation of the conifold. In \cite{buchel} instead this procedure has
been reversed, trying to make predictions on the field theory using string theory
computations. It was suggested that a non-extremal generalization of the KT solution may
have a regular Schwarzschild horizon ``cloaking'' the naked singularity. 
Non-extremal supergravity solutions translate in field theories at finite temperature \cite{finitet}. In this case, the dual field theory interpretation of this has been then supposed to be the restoration of chiral symmetry
at a finite temperature $T_C$ \cite{buchel}.

From the supergravity point of view, we are then in a particular  situation: at zero
temperature the theory exhibits chiral symmetry breaking \cite{KS}, while for temperature
above $T_C$ we have chiral symmetry restoration. So, the regular Schwarzschild horizon
should appear only for some finite Hawking temperature. As noticed in
\cite{buchel,non-extremal,restoration}, one implication is that, at temperatures below
$T_C$, there should be non-extremal generalizations of the Klebanov-Strassler solution
which are free of horizons, just like the extremal solutions. Particularly, the analysis
of \cite{buchel,non-extremal,restoration} suggests that gravitational backgrounds which
exhibit regular Schwarzschild horizon only above some critical non-extremality, should be
quite generic for backgrounds dual to gauge theories which undergo finite temperature
symmetry breaking phase transition. Nevertheless, they are unusual and, to our knowledge,
new from the supergravity point of view.

We argue that supergravity backgrounds of that type have to involve the enhan\c{c}on
mechanism in non-extremal backgrounds, so that a combination of the enhan\c{c}on shell
effect and of the horizon of the black hole, determines the appearance of regular
Schwarzschild horizons only in particular range of the parameters of the solutions. We
leave anyway this issue for future investigation.

We expect similar infrared phenomena to take place in the fractional D0-, D1-, and
D2-brane cases. Easily generalizing  the observation and the computation of \cite{KS,PZ}
we can say that there are two natural ways of smoothing out the singularity at the apex of
the conifold: one can substitute the apex by a $(6-p)$ non trivial cycle (deformation) or
by a 2-cycle (resolution). We expect anyway that the resolution still displays a naked
singularity (as it happens for the $p=3$ case \cite{PZ}), that can be excised by the
enhan\c{c}on mechanism. Indeed, as pointed out in \cite{KS} for the $p=3$ case, the source
of the singularity can be traced back to the infinite energy of the $F_{6-p}$ (magnetic)
field. At all radii there are $M$ units of flux of $F_{6-p}$ along the $(6-p)$ cycle, and
when that cycle shrinks to zero size, this causes $F_{6-p}^2$ to diverge. Since resolution
does not affect the $(6-p)$-cycle, we can think naively that, if we resolve the conifold,
the solution remains singular. However the enhan\c{c}on mechanism prevents $F_{6-p}^2$
from diverging, so it's possible, as we will show, to have a non-singular solution also in
the case of resolution. Instead, deforming the conifold does mean exactly that the $(6-p)$
cycle does not shrink, so the singularity can be avoided (as it happens in the deformed
conifold solution of \cite{KS}).

In this paper we will analyze the enhan\c{c}on mechanism on the conifold with the aim of
checking if it's consistent with the physical expectations discussed above. We will
perform this analysis using exclusively supergravity tools. However the enhan\c{c}on
mechanism in the presence of fractional branes occurs also in the case of orbifold
backgrounds \cite{pol,myers,bertolini,frau}. These are more tractable examples since we
have an exact stringy description of these geometries. We will then concentrate on those
backgrounds, where we can compare the supergravity solutions with some new insights we can
have from the boundary-state formalism.  Thus, we will construct, in Appendix A, the
boundary state appropriate to describe fractional branes.

The analysis of how supergravity results perfectly match with stringy computations in the
orbifold case, corroborates, in our opinion, the interpretation we propose also for
conifold backgrounds. In particular, in the orbifold case, we will see that the NS-NS
twisted sector contributes to the definition of what has to be interpreted as the
effective mass of a fractional D$p$-brane. This interpretation will be supported by
different calculations, involving the boundary-state techniques as well as the
supergravity description of branes. From both these types of computations, we'll find
that, as in the case of branes wrapped around $K3$ \cite{pol,myers}, the volume of the
internal manifold plays a fundamental role. In particular we will then show that when the
volume reaches a critical value, the fractional branes become massless.

Finally, we'll find that, in both conifold and orbifold backgrounds, in order to have a
consistent enhan\c{c}on mechanism, the shell of branes defining the enhan\c{c}on locus is
made up by fractional branes only and that the presence of some regular D$p$-brane in the
origin is necessary.

\section{The enhan\c{c}on on the conifold}

Type IIA and type IIB fractional D$p$-brane solutions in conifold backgrounds have been
extensively studied in \cite{herzog}. They are warped product of $\mathbb{R}^{1,p}$ flat
space-time directions and a Ricci flat, ($9-p$)-dimensional cone $\mathcal{C}_{9-p}$.
Since the brane are space filling, the warp-factor depends only on the radial coordinate
of the cone. Because the cone is Ricci flat, the base of the cone is an
($8-p$)-dimensional Einstein manifold $X_{8-p}$ with metric $h_{ij}$. It's assumed this
Einstein manifold has a harmonic 2-form $\omega_2$, so that wrapping a D$(p+2)$-brane
around the 2-cycle corresponding to $\omega_2$, and letting the remaining $p+1$ dimensions
fill $\mathbb{R}^{1,p}$, gives origin to a fractional D$p$-brane.\\ We concentrate now on
the fractional D$p$-branes, with $p<3$.

\subsection{Fractional D$p$-branes, $p<3$}

As a generic example and to see what it is meant by warped product we reproduce the D0
fractional brane solution found in \cite{herzog}, but our considerations will be easily
generalizable for every $p$:
\begin{eqnarray}
\no \mbox{e}^{\Phi} &=& H(r)^{3/4}  \\
ds_E^2 &=& g_s^{1/2}\ \Big[ -H^{-7/8}dt^2+H^{1/8}(dr^2\ +\ r^2h_{ij}dx^i\ dx^j)\Big]
\end{eqnarray}
The non-zero R-R field strengths are:
\begin{eqnarray}
\no F_2  &=& dt\wedge dH^{-1}\\
\tilde{F}_4 &=& \frac{H^{-1}}{r^4}~Q~dt\wedge dr\wedge\omega_2
\end {eqnarray}
where $\tilde{F}_4\ =\ F_4- C_1\wedge H_3$, $Q\sim   g_sM$, $M$ is the number of fractional D$p$-branes
and $g_s$ is the string coupling constant. \\
We want now to express, for more generality, the NS-NS two form field ($B_2$) and the warp
factor $H$ for $p$ generic ($p<3$):
\begin{eqnarray}
\no B_2 &=& \frac{Q}{r^{3-p}}\frac{\omega_2}{p-3}\\
\label{Hsol} H(r) &=& \frac{\rho}{r^{7-p}}\ -\ \frac{Q^2}{(3-p)(10-2p)r^{10-2p}}
\end{eqnarray}
where as usual $\rho\sim g_s N$ and $N$ is the number of ordinary D$p$-branes. The warp
factor has to satisfy the following equation of motion (it comes from that for the R-R
field strength $F_2$):
\begin{equation}
\label{H} \frac{d}{dr}\Big(r^{8-p}\frac{d}{dr}H(r)\Big)=-\frac{Q^2}{r^{4-p}}
\end{equation}

This is the Poisson equation for the $H$-potential. It has been integrated in (\ref{Hsol})
with the boundary condition that $H$ approaches zero as $r\to\infty$, so this solution is
valid for small $r$ (near-horizon limit) and consequently we are studying the gauge theory
(according to the gauge/gravity dual principle) in the infrared. \\ Note also that the
$r$-dependence of the source term seems to indicate a spatial extension of the source in
the transverse directions. Moreover, from the precise form of the dependence, it's
possible to argue that it spreads over ($4-p$) directions of the base of the cone.
\\ Moreover, as already pointed out by the authors of \cite{herzog}, such solutions possess a
naked singularity at $r=r_0$, in the IR. This singularity is of ``repulson'' type because
before reaching it, the derivative of $G_{00}$ changes sign at $r_*>r_0$ (where
$H'(r_*)=0$) and consequently the gravitational force changes its sign.

We want now study the possibility that an excision process takes place, as it has been
shown to happen in \cite{myers} in a different context. There, it has been studied this
mechanism for branes wrapped around the $K3$ surface, but we think that also in the
present case it happens something very similar, giving rise to the enhan\c{c}on. \\ Then,
in order to avoid the singularity, we can imagine that the D$p$-branes are not in the
origin, but they distribute on a surface at some $r=r_i\geq r_*$. For generic $p<3$ we can
try to calculate the stress-energy tensor of these source branes, interpreted as the
discontinuity in the extrinsic curvature (we are making the calculation in the Einstein
frame). Following \cite{myers} we define unit normal vectors
$$n_{\pm}\ =\ \mp\frac{1}{\sqrt{G_{rr}}}\frac{\partial}{\partial r}.$$
The extrinsic curvature of the junction surface, for the two regions ($r>r_i$ and $r<r_i$)
is $$ K^{\pm}\ =\ \frac{1}{2} n_{\pm}^c\partial_{c}G_{AB}\ =\
\mp\frac{1}{2\sqrt{G_{rr}}}\frac{\partial G_{AB}}{\partial_r} $$ The discontinuity in the
extrinsic curvature is then defined as $\gamma_{AB}\ =\ K_{AB}^{+}+K_{AB}^-$. Finally the
stress-energy tensor supported at the junction is simply\footnote{Throughout this paper,
the indices $A$, $B$ run from 0 to 9, $\mu$ and $\nu$ denote the world-volume directions
of the brane, while the indices $i,\ j$ denote the transverse ones}:
\begin{equation}
S_{AB}\ =\ \frac{1}{\kappa^2}\Big( \gamma_{AB}-G_{AB}\gamma^C_C\Big)
\end{equation}where $\kappa$ is the gravitational coupling constant.

Thus, if we think that in the interior region ($r<r_i$) the space is flat,
with easy calculations we obtain:
\begin{eqnarray}
S_{\mu\nu} &=& \frac{1}{2\kappa^2}\frac{1}{\sqrt{G_{rr}}}\Big(\frac{H'}{H}\Big)G_{\mu\nu} \\
\no S_{ij} &=& 0
\end{eqnarray}
According to the second line, we see that there is no stress in the directions transverse
to the branes, as it has to be since the branes in consideration are BPS.\\
From the first line we can argue that for $r_i=r_*$ the constituent branes become
massless. To get more information about this configuration we can now expand the result
for large values of the incision radius $r_i$. The coefficients of the metric components
in the longitudinal directions ($\mu,\ \nu$) can be interpreted as proportional to the
effective tension of the brane. So we see that:
\begin{equation}
\tau (r_i)=\frac{1}{2\kappa^2}\frac{H'}{H}=\frac{1}{2\kappa^2}\frac{ (p-7) +
\frac{Q^2}{(3-p)\rho r_i^{3-p}}}{1-\frac{Q^2}{\rho (3-p)(10-2p)r_i^{3-p}}}\frac{1}{r_i}\
\sim\
\frac{1}{2\kappa^2}\Bigg(\frac{p-7}{r_i}+\frac{Q^2}{(10-2p)\rho}\frac{1}{r_i^{4-p}}\Bigg)
\end{equation}
We can notice that this expression seems to indicate a configuration that does not
reproduce the behaviour of the source term in the equation of motion (\ref{H}).\\ Thus we
now imagine a slightly different configuration: we think that all the $N$ ordinary branes
are in the origin and the fractional branes alone form a shell at $r=r_i$. We can now
repeat the computation of the stress-energy tensor for this configuration and we get
always a vanishing contribution in the transverse directions but in the longitudinal ones
we have the following result:
\begin{equation}
\tau (r_i)\ \sim\ \frac{1}{2\kappa^2}\frac{M}{N}\frac{Q}{(10-2p)}\frac{1}{r_i^{4-p}}
\end{equation}
This result is now compatible with the source term which appear in (\ref{H}). This
expression for the effective tension is due
to the presence of $M$ fractional branes and it contains the small parameter $M/N$ that is typical for these expansions.\\
We get then the following solution:
\begin{eqnarray}
H_p(r) &=& h_s(r)+\Theta(r-r_i)[h_f(r)-h_f(r_i)]\\
\no H'_p(r) &=& h'_s(r)+\Theta(r-r_i)h'_f(r)\\
\no H''_p(r) &=& h''_s(r)+\Theta(r-r_i)h''_f(r)+\delta(r-r_i)h'_f(r)
\end{eqnarray}
where $$h_s(r)=\frac{\rho}{r^{7-p}}\ \ \ \mbox{and}\ \ \
h_f(r)=-\frac{Q^2}{(p-3)(10-2p)}\frac{1}{r^{10-2p}}$$ For the sake of clarity, we report
here the modified ansatz and the equations of motion for the case of a D$0$ fractional
brane, but everything continues to be easily generalizable to other $p$:
\begin{eqnarray}
&&\no F_2=dt\wedge dH^{-1}\\
&&\no B_2=\frac{Q}{p-3}\frac{\Theta(r-r_i)}{r^{3-p}}\ \omega_2\\
&&\no \tilde{F}_4=H^{-1}\frac{Q\ \Theta(r-r_i)}{r^{4-p}} dt\wedge \ dr\wedge\omega_2\\
&&\no d(\mbox{e}^{\phi/2}\ *\tilde{F}_4)=Q\delta(r-r_i)\\
&&\label{theta} d(\mbox{e}^{\frac{3}{2}\phi}\ *F_2)=g_s\mbox{e}^{\phi/2}H_3\wedge
*\tilde{F}_4\ =\
-\frac{Q^2}{r^{4-p}}\Theta(r-r_i)-\frac{Q^2}{p-3}\frac{\delta(r-r_i)}{r^{3-p}}
\end{eqnarray}
With the new solutions all the singular terms (see in particular the last equation) fit
perfectly. They are those which give rise to the junction conditions. One could also study
the dilaton equation of motion to see it has a discontinuity at $r=r_i$ (see \cite{myers})
but it is not relevant here.

Thus the scenario seems to be the following one. We have D$(p+2)$-branes at $r=r_i$ that
are wrapped around the 2-cycle which is Poincar\`e dual to $\omega_2$. According to the
form of their effective tension, it seems these objects extend in ($4-p$) of the
directions of the base of the cone. Moreover, these have a non trivial coupling with the
$B_2$ field, that becomes a source for fractional D$p$-branes potential for all $r\geq
r_i$ (notice the form of the source term in (\ref{theta})). \\ This is anyway suggested
also by the running of the fluxes:
\begin{equation} \Phi\ =\ \int_{\mathcal{C}_{9-p}} d(\mbox{e}^{\frac{3}{2}\phi}*F_2)\ =
\rho+\Big(\frac{1}{3-p}\frac{Q^2}{r^{3-p}}-\phi_*\Big)\Theta(r-r_*)
\end{equation}
where $\phi_*=\frac{1}{3-p}\frac{Q^2}{r_*^{3-p}}$. This behaviour seems indeed to indicate
that the fractional branes are present for all $r\geq r_i$ and that, at fixed $r$, they
extend in the ($4-p$) directions of the base of the cone where the shell at $r=r_i$
spreads.

In our discussion we have never fixed $r_i$, but we found a lower bound for it:
$$r_*^3=\frac{Q^2}{(7-p)(3-p)\rho},$$ where we argued the branes become tensionless. This
is very reminiscent of the enhan\c{c}on locus \cite{pol}, where we know the branes become
effectively massless. We think indeed it is now working the same mechanism, that seems
typical for fractional D-branes in various background \cite{bertolini,frau,pol}. In this
case it would be natural to identify the enhan\c{c}on radius $r_e$ with $r_*$. It would be
nice to find some interpretation of this with the boundary states or other conformal
techniques, but, as already emphasized in the introduction, we don't have these tools at
our disposal in this case.
 So in the following section we apply the same techniques used
in the present one to other backgrounds which do have a conformal description and we'll
see how the supergravity computations match with the stringy ones. We'll get therefore
convinced of the right interpretation of the supergravity results.

Before that, we consider now the $p=3$ case. This case is indeed particularly relevant in
the context of the gauge-gravity duality \cite{conifold} and the formulae are slightly
different from the other cases.

\subsection{Fractional D$3$-brane}

In the context of gauge/gravity duality the basic picture common to all RG problems is
that the radial coordinate defines the RG scale of the field theory, hence the scale
dependence of the couplings may be read off from the radial dependence of the
corresponding supergravity fields. Since the RG flows of couplings in physically relevant
gauge theories are logarithmic, an important problem is to find gravity duals of
logarithmic flows. Thus the importance of the case of the fractional D3-branes. Indeed the
solution describing a collection of $N$ D3-branes and $M$ fractional D3-branes on the
conifold was constructed in \cite{KT} (where the base of the cone is the $T^{1,1}$
manifold), finding a logarithmic radial dependence of certain supergravity fields.

The gauge dual of this background is an $N=1$ supersymmetric non-conformal gauge theory.
Besides the conical case we are considering now, other supergravity solutions representing
such configurations have been constructed \cite{KS,PZ} also for the deformed conifold
\cite{KS} and the resolved conifold \cite{PZ}. The two solutions, for large values of the
radial coordinate, (in the UV from the dual gauge theory perspective), coincide with the
conifold one we examine. However, the small-distance (IR) behaviour is different in each
case. In particular, the conifold and the resolved conifold have naked singularities at
finite values of the radial coordinate, while the deformed conifold solution is regular.

Let us now perform the excision calculation for the conifold solution. We want to notice
that also in the case of the resolved conifold \cite{PZ} the metric can be written, in
term of a different radial variable $\rho$ \cite{PZ}, in a way which is very similar to
the standard conifold case (at least in the IR, where the solution exhibits its
singularity). The
excision mechanism can therefore be used to avoid the singularity also in that case.\\
Now we illustrate it explicitly.

The metric is a warped product
\begin{equation}
ds^2~ =~ H(r)^{-1/2}\eta_{\mu\nu}dx^{\mu}dx^{\nu}~+~H(r)^{1/2}(dr^2+r^2h_{ij}dx^idx^j)
\end{equation}
and the equation of motion (or the Bianchi identity) for the self-dual field-strength
$\tilde{F}_5$ reads
\begin{equation}
\label{F5} d\tilde{F}_5~=~ H_3\wedge F_3\ \ \ \ \ \ \ \ \mbox{where}\ \ \ \ \ \ \ \
\tilde{F}_5~=~ dC_4~+~B_2\wedge F_3.
\end{equation}
If we make the usual ansatz for fractional D3-branes
\begin{eqnarray}
\no F_3 &=& Q\omega_3\\
\tilde{F}_5 &=& d^4x\wedge dH^{-1}~+~*(d^4x\wedge dH^{-1})\\
\no H_3 &=& \frac{Q}{r} dr\wedge\omega_2
\end{eqnarray}
equation (\ref{F5}) becomes
\begin{equation}
(r^5 H')'~=~ \frac{Q^2}{r}.
\end{equation}
We can integrate it with the appropriate boundary condition to study the near-horizon
limit (IR dynamic of the dual gauge theory), and obtain:
\begin{equation}
H~=~\frac{\rho}{r^4}~+~ Q^2\Big( \frac{\ln \frac{r}{R}}{4r^4}~+~\frac{1}{16 r^4}\Big)
\end{equation}
where $R$ is a scale of the same order of $r_*$. If we repeat for this case the excision
calculation, with the usual procedure, we get a stress-energy tensor at the incision
radius $r_i$:
\begin{eqnarray}
\label{long3}S_{\mu\nu} &=& \frac{1}{2\kappa^2}\Big(\frac{1}{2\sqrt{G_{rr}}}\frac{Q^2/4r^5}{H(r)}G_{\mu\nu}\Big)\\
\no S_{ij} &=& 0
\end{eqnarray}
where we have already considered the ordinary branes in the interior region of the shell
and the fractional ones forming the shell.

All the considerations we made for the other $p$ continue to be valid. In particular,
notice that the right expansion for (\ref{long3}) seems to be for $r\approx R$, (where $R$
is a scale such that $R\approx r_*$\footnote{also in this case $r_*$ is defined as
$H'(r_*)=0$}). We obtain (for $M\ll N$):
\begin{equation}
S_{\mu\nu}\approx \frac{1}{2\kappa^2}\frac{M}{N}\frac{Q}{4}\frac{1}{r_i}.
\end{equation}
This is again consistent with the equations of motion and it seems to suggest that the
shell is located precisely at $r_*$, the radius we argued has to be identified with the
enhan\c{c}on radius.

We have already mentioned that the deformed conifold solution is regular. In that case the
mechanism that removes the singularity is related to the breaking of the chiral symmetry
in the dual $SU(N)\times SU(N+M)$ gauge theory. The $\mathbb{Z}_{2M}$ chiral symmetry,
which may be approximated by $U(1)$ for large $M$, is broken to $\mathbb{Z}_2$ by the
deformation of the conifold \cite{KS}.

In \cite{buchel} a different mechanism for resolving the naked singularity was proposed.
It was suggested that a non-extremal generalization of the solution found in \cite{KT} may
have a regular Schwarzschild horizon ``cloaking'' the naked singularity. The dual field
theory interpretation of this would be the restoration of chiral symmetry at a finite
temperature $T_C$ \cite{buchel}. In \cite{non-extremal} the right ansatz and the system of
second order radial differential equations necessary to study such non-extremal solution
have been constructed. Finally, in \cite{restoration}, smooth solutions to this system
have been found in a perturbation theory that is valid when the Hawking temperature of the
horizon is very high.\\ Clearly it would be very interesting to study this theory as a
function of the temperature and to identify the phase transition at $T=T_C$. We argue that
it's possible that the excision mechanism is valid for studying the physics of the
enhan\c{c}on at finite temperature, as it seems to happen in orbifold backgrounds
\cite{pol,myers,myers2}. Notice indeed that this solution does not break chiral symmetry.
Particularly, it would be interesting to investigate the possibility this mechanism is
related to the non-extremal solution at $T=T_C$, where
chiral symmetry is restored, but we leave the study of this possibility for future work.\\
Let us now concentrate on the orbifold background.

\section{The enhan\c{c}on on the orbifold}

\subsection{Comparison with the conifold}

Now we want to examine the supergravity solutions for fractional D$p$-branes in an
orbifold background
\begin{equation}
\label{bg} \mathbb{R}^{1,5}\times T^4/\mathbb{Z}_2,
\end{equation}
where $T^4$ is the four-dimensional torus. For this purpose we mainly refer to
\cite{frau}, where the solutions have been written down explicitly for $p=0\ ,2$ in the
context of the IIA theory. We report here the solutions, integrating the equation for the
warp factor with the boundary condition that $H$ approaches zero as $r\to\infty$, as we
did in the previous paragraph:
\begin{eqnarray}
\no ds^2 &=& H^{-(3-p)/4}\eta_{\mu\nu}dx^{\mu}dx^{\nu}+H^{(p+1)/4}(dr^2+r^2d\Omega^{4-p})\\
\no \mbox{e}^{\varphi} &=& H^{(1-p)/4}\ \ \ \ \ ,\ \ \ \ \ \mbox{e}^{\eta_a}\ =\ H^{1/4}\\
\label{orbsol} H &=& \frac{1}{2}\frac{Q_p}{r^{3-p}}-\frac{1}{2}\frac{\mathcal{V}}{(2\pi)^4\ap^2}\frac{Q_p^2}{r^{6-2p}}\\
\no \tilde{b} &=& -\frac{1}{\sqrt{2}}\frac{\mathcal{V}^{1/2}}{4\pi^2\ap}\frac{Q_p}{r^{3-p}}
\end{eqnarray}
where $$ Q_p\ =\ \frac{2\sqrt{2}MT_p\kappa_{orb}}{(3-p)\Omega_{4-p}\mathcal{V}^{1/2}}, $$
$M$ is the number of fractional branes, $\kappa_{orb}=\sqrt{2}\kappa/\mathcal{V}^{1/2}$,
$\ \Omega_{4-p}$ is the area of a unit ($4-p$) dimensional sphere, $r$ is the radial
coordinate in the ($5-p$)-dimensional space transverse to the brane, $\varphi$ is the
six-dimensional dilaton, $\tilde{b}$ is the twisted NS-NS scalar and $\eta_a$ are the four
untwisted NS-NS scalars:
\begin{equation}
\label{orbsol2} G_{aa}\ =\ \mbox{e}^{2\eta_a}\ \ \ \ \ \mbox{     }\ a=6,\ldots, 9
\end{equation}
and finally $G$ is the ten dimensional metric in the string-frame. We don't report here
explicitly the form of the R-R twisted and untwisted potentials because they are not
relevant in our discussion.\\
As already pointed out in \cite{frau}, this solution presents a naked singularity, that is
of repulson type because it's located at $r_0<r_*$, ($r_*$ is defined so that
$H'(r_*)=0$).\\
From what we know from the enhan\c{c}on and from what we have learned in the conifold
case, we are now naturally led to imagine that also in this case the fractional branes
form a shell at $r=r_i \geq r_*$. Analogously to the computation made in the previous
paragraph, we can then calculate the stress energy tensor for this configuration at the
junction surface. We get:
\begin{eqnarray}
\label{shell} S_{\mu\nu}(r_i) &=&
\frac{1}{2\kappa^2_{orb}}\frac{1}{\sqrt{G_{rr}}}\Big(\frac{H'}{H}\Big)G_{\mu\nu},\\
\no S_{ij} &=& 0
\end{eqnarray}
The vanishing of the stress-energy in the transverse directions again indicates that the
system is BPS and because the stress energy tensor in the longitudinal directions is
proportional to $H'$, it's again natural to argue that at $r_*$ the constituent branes
become massless. The advantage we have in this case, it's that this interpretation is now
strongly supported by a direct probe calculation (made in \cite{frau}). Indeed in the
orbifold background we are considering, we have a stringy description that enables us to
construct the world-volume action describing the interaction of the D-brane with the
fields of the spectrum of the theory. For the background (\ref{bg}), this has been written
in \cite{frau} and the part of interest for us (obtained expanding the DBI part of the
world-volume action in the velocities and keeping only the lowest order terms), is:
\begin{equation*}
  - ~ \frac{T_p}{\sqrt{2\mathcal{V}}k_{orb}}\int d\zeta^{p+1} \delta_{ij} \frac{\partial
x^i}{\partial\tau}~\frac{\partial x^j}{\partial\tau}\Big( 1 +
\frac{\sqrt{2\mathcal{V}}}{2\pi^2\ap}\tilde{b}\Big)
\end{equation*}
We see that the interaction term of the brane with the NS-NS twisted field seems to add a
contribution to the definition of the tension of the brane. \\
We can interpret it as if both the NS-NS sector (with a constant contribution) and the
NS-NS twisted one (with an $r$-dependent term) contribute to the definition of the
``running'' tension of the brane. Then, it is easily shown \cite{frau} that the
enhan\c{c}on radius $r_e$, defined as the radius where the running tension of the brane
goes to zero, coincides with $r_*$.

Comforted by these results, we consider now the stress-energy tensor (\ref{shell}) and, as
in the conifold case, we expand it for large $r_i$. We find again a configuration not
consistent with the equations of motion.

So we can now imagine of making analogous considerations for a configuration in which we
add $N$ ordinary D$p$-branes in the origin (with $N\gg M$). We assume furthermore that the
$M$ fractional D$p$ branes alone form the composite shell. In this case, performing an
excision calculation,we get:
\begin{equation}
S_{\mu\nu}\ \sim\ \frac{1}{2
\kappa^2}\frac{M}{N}\frac{(3-p)Q_0\mathcal{V}}{(2\pi)^4\ap}\frac{1}{r_i^{4-p}}
\end{equation}
and this is consistent with the equations of motions.\\ Then we notice that, in order to
have a consistent enan\c{c}hon mechanism, excision computations tell us we have to insert
ordinary branes in the interior region, both in the conifold and in the orbifold case.\\
Moreover it seems again the branes smear in ($4-p$) directions. Being the spacetime
(\ref{bg}) six-dimensional, these directions are naturally interpreted as the angular
space-time dimensions transverse to the brane. However it's interesting to note this is
the same number ($4-p$) that appears in the conifold case, where we have no this
interpretation.

Finally, we want to emphasize that this solution shares a particular feature with the
enhan\c{c}on solution found in \cite{pol}, that is the $r$-dependent compactification volume.\\
Indeed, from (\ref{orbsol}-\ref{orbsol2}), we see that the volume of the orbifold, as
measured in the string-frame, is:
\begin{equation}
\frac{V(r)}{2} =\ \mathcal{V}\prod_a \mbox{e}^{\eta_a}\ =\ \mathcal{V}H(r)
\end{equation}
where the factor 2 is typical for orbifold compactifications. We see it depends on $r$ and
that at $r=r_e$ it takes the value:
\begin{equation}
\label{vol}
V(r_e)\ =\ (\sqrt{2\ap}\pi)^4
\end{equation}

\subsection{Effective mass and the twisted sectors}
\label{section}

We want now to use the description of fractional branes by means of the boundary state
(see Appendix A). In this paragraph, for the sake of clarity, we'll consider branes that
don't extend in the compact dimensions. The aim of this analysis is to extract information
about the vanishing of the tension that seems to occur at a particular value of the
space-time radial coordinate ($r=r_e$) and for the value of the compactification volume
(\ref{vol}).

To get information about the mass of a brane, it is sufficient to study  the long-distance
behavior of the $p$-brane solution. This can be obtained directly from the boundary state.
Indeed, as shown in \cite{classical}, it is possible to derive the most singular term in
$k$ (the transverse impulse) of the classical solution by projecting the boundary state
onto the level-one states with the closed string propagator inserted in between:
\begin{equation}
\label{corrente} J(k)\ =\ \langle0;0|V^{NS-NS}(k;z,\bar{z})P|B\rangle_{NS}
\end{equation}
where P is the closed string propagator, $V^{NS-NS}(k;z,\bar{z})$ is the vertex operator
producing the level-one states and $|B\rangle$ is the boundary state (see Appendix A). The
symmetric irreducible component of the current gives the long-distance behavior of the
graviton, from which we can easily extract the (density of) mass $M$ of the brane. Indeed,
when saturated with the graviton polarization ($h^{\mu\nu}$), we expect it has the
following form: $$\mbox{Tr}(h ~ J) ~ = ~ -M ~ \frac{1}{k^2} ~ V_{p+1} ~ \mbox{Tr}(h\cdot
L)
$$ where the longitudinal matrix $L$ is the projector on the world-volume directions of
the brane.

\begin{figure}[t]
  \centering
  \epsfig{figure=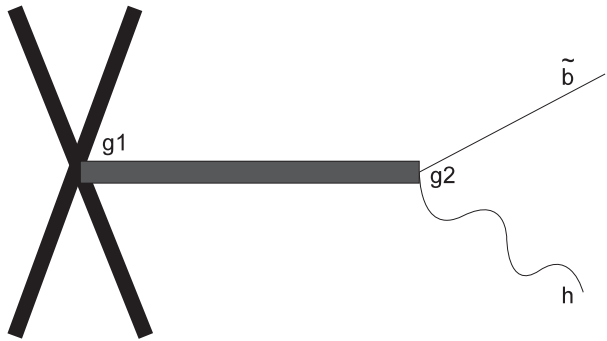,width=8cm}
\caption{The two-point diagram}
  \label{figure1}
\end{figure}

We can follow the same procedure for the fractional branes. The difference is that the
normalization constant of the boundary state is now different in the NS-NS untwisted
sector. We then obtain:
\begin{equation}
\label{massu} \mbox{Tr}(h ~ J) ~ = ~ - \frac{T_p}{\sqrt{2V}} ~ \frac{1}{k^2} ~ V_{p+1} ~
\mbox{Tr}(h\cdot L),\ \ \ \ \ \mbox{and then}\ \ \ \ \ M ~ = ~ \frac{T_p}{\sqrt{2V}}
\end{equation}
But in the case of fractional branes we think this is not the end of the story. Indeed we
have that, even if the linear coupling between the NS-NS twisted sector and the graviton
is vanishing, this sector provides a two-point amplitude between the graviton and the
NS-NS twisted scalar, which is different from zero (see figure \ref{figure1}). This
amplitude is given by \begin{equation}\label{ampl}\langle 0;0|V^{NS-NS~T}(k;z_2,\bar{z}_2)
V^{NS-NS}(k;z_1,\bar{z}_1)P|B\rangle_{NS}.\end{equation} The main point is that the NS-NS
twisted sector, even if not linearly, couples to the graviton. This suggests it can
participate in some way at the definition of mass, in agreement with the probe
calculation. In \cite{merlsab} the amplitude (\ref{ampl}) has been calculated, obtaining:
\begin{equation}
\label{twopoint} A(h,\tilde{b})\ =\ - N_{T,p} ~ V_{p+1} ~ \mbox{Tr}(h\cdot L) ~ \tilde{b}
\end{equation}
where $N_{T,p}$ is defined in appendix A to be:
\begin{equation*}
  N_{T,p}\ =\ \frac{\sqrt{2}T_p}{(4\pi^2\ap)}
\end{equation*}
We can now extract from (\ref{twopoint}) the contribution (we call it $\mathcal{M}$) of
the NS-NS twisted sector to the (density of) mass of the brane. We argue indeed that
(\ref{twopoint}) can be brought to the form:
\begin{equation}
 \label{g} A(h,\tilde{b})\ =\ g1 \cdot g2 \cdot \mathcal{M} ~ \frac{1}{k^2} ~ V_{p+1} ~ \mbox{Tr}(h\cdot L).
\end{equation}
We have then to identify $g1$ and $g2$, (see figure \ref{figure1}). At this purpose it's
useful to note that this amplitude is different from zero because of the presence of the
NS-NS twisted scalar in the closed string propagator. We can then identify $g1$ with its
linear coupling with the brane and $g2$ with the strength of the coupling of the graviton
with two twisted NS-NS scalars. Thus we have \cite{merlsab}:
\begin{equation}
\label{1} g1 ~ = ~ -\sqrt{2}N_{T,p}\ \ \ \ \ \mbox{and}\ \ \ \ \ g2 ~ = ~ k_6
\end{equation}
Moreover, it's easy to see from (\ref{orbsol}) that the field $\tilde{b}$, in the
impulse-space, has the following form:
\begin{equation}
\label{TF} \tilde{b}(p)\ =\ -(\sqrt{2}N_{T,p}\kappa_6)\frac{1}{k^2}
\end{equation}
Finally, from (\ref{twopoint}-\ref{g}-\ref{1}-\ref{TF}), we argue that:
\begin{equation}
 \label{massT} \mathcal{M} ~ = ~ N_{T,p}
\end{equation}
If we then consider both the contributes (\ref{massu}-\ref{massT}) we get a total density
of effective mass:
\begin{equation}
M\ =\ \frac{T_p}{\sqrt{2V}}-N_{T,p}
\end{equation}
It's now easy to see, from the definition of $T_p$ and $N_{T,p}$ (see appendix A), it
vanishes precisely at the value (\ref{vol}) of the internal volume.

We can then conclude that the vanishing of the (effective) mass of a D$p$-brane is
strictly linked to the presence of twisted sectors and, particularly, to a precise value of the
compactification volume. These ingredients are all encoded in the boundary
state. In principle one could then expect to obtain this result by direct computations
involving only branes and their description via the boundary state. The straightforward
calculus we can make it's the tree-level closed string amplitude:
\begin{equation}
  \langle B|P|B\rangle
\end{equation}
We make this computation and comment it in the appendix A, where we also construct the
generic boundary state describing fractional wrapped branes.

\section{Conclusions}

In this paper we have shown how the enhan\c{c}on mechanism can resolve the repulson
singularities both in conifold and orbifold backgrounds.\\
In the case of the conifold there are no conformal techniques at our disposal but,
uniquely by means of supergravity tools, we have shown how the enhan\c{c}on mechanism
seems to take place and resolve the singularities. We have then studied the enhan\c{c}on
mechanism by means of the same supergravity tools in the orbifold background, where we
have in hand also conformal techniques. We have found a perfect matching between the two
computations, getting therefore persuaded of the correctness of the supergravity
computations and of their interpretation in the case of the conifold.

In the conifold background we have examined the solutions for D$p$-branes with $p\leq 3$.
We have shown how, for all those cases, the excision computation is consistent with the
enhan\c{c}on mechanism. We have paid particular attention to the case $p=3$, because of
his relevance for the dual gauge theory. In that case, this mechanism is good at avoiding
the singularity also in the case of the resolved conifold \cite{PZ} and, particularly, we
expect it to be good at describing the non extremal solutions at temperatures above chiral
symmetry breaking temperature ($T_C$) in the gauge theory. We leave this issue for future
investigation.

In the orbifold background, we have emphasized the role of the twisted NS-NS sector in
contributing to the definition of an effective mass for a fractional brane. We have shown
how the vanishing of this effective mass is related to the volume of the internal
manifold. This relation is supported in the Appendix by the computation of the three level
closed string amplitude between two D$p$-branes. This calculation is based on the
open/closed string duality. We show that a convenient choice of the sign of the orbifold
projection on the ground state of the open string leads to determine the value of the
internal volume at which the brane becomes effectively massless. The opposite choice leads
instead to suppose the existence of non-BPS ``truncated'' D$p$-branes that are stable for
particular values of their moduli. We make that calculation for generic wrapped branes.

Finally, we have shown that, in both the conifold and orbifold backgrounds, consistency of
the excision process requires that the interior space is non-empty.

\section*{Acknowledgement}

I would like to thank A. Lerda and G. Sabella for very useful discussions. I'm also
grateful to M. Bertolini and M. Bill\`o for enlightening comments on a preliminar version
of the paper.

\newpage

\appendix{\section{Construction of the boundary state for wrapped fractional branes}}

We want to derive the structure and the normalizations of the boundary state in the
orbifold background
\begin{equation}
\label{orbifold} \mathbb{R}^{1,5}\times T^4/\mathbb{Z}_2
\end{equation}
where $\mathbb{Z}_2$ is the reflection parity that changes the sign to the four
coordinates of the torus $T^4$. The D$p$-brane is intended to have $r+1$ longitudinal
space-time directions and $s$ that extend in the compact manifold ($p=r+s$). In an
orbifold background like (\ref{orbifold}) the boundary state has four different components
which correspond to the (usual) NS-NS and R-R untwisted sectors and to the NS-NS and R-R
twisted sectors \cite{9906242,9910109,9912157}.

In each sector of the theory we can construct the boundary state:
\begin{equation}
 |B,k,\eta\rangle  =
  \esp^{i\theta} \mbox{ exp}
	\Bigg(\sum_{l>0}\Big[\frac{1}{l}\alpha_{-l}^{\mu}S_{\mu\nu}\tilde{\alpha}^{\nu}_{\-l}
					\Big]
		~ + ~ i\eta \sum_{m>0}\Big[\psi_{-m}^{\mu}S_{\mu\nu}\tilde{\psi}^{\nu}_{\-m}
					\Big]
	\Bigg)|B,k,\eta\rangle^{(0)}
\end{equation}
where $l$ and $m$ are integer or half-integer depending on the sector, $k$ denotes the
momentum of the ground state and $\theta$ is a phase equal to $\pi$ in the untwisted R-R
and twisted NS-NS sectors, to $3\pi/2$ in the untwisted NS-NS sector and to $0$ in the
twisted R-R sectors. The parameter $\eta=\pm 1$ describes the two different spin
structures \cite{Pol,Cal}, and the matrix $S$ encodes the boundary conditions of the
D$p$-brane which we shall always take to be diagonal. For an exhaustive
review on this subject see \cite{anto}\\
We omit always the ghost and superghost part of the boundary states, which are as in the
usual case \cite{9707068,9802088}. Finally, we want to stress that in the two untwisted
sectors, the ground state is in addition characterized by a winding number $n_i$ for each
compact
direction that it's tangential to the world-volume of the brane. \\
 In order to obtain a
localized D$p$-brane (say in $y=0$), we have to take the Fourier transform of the above
boundary state, where we integrate over the directions transverse to the brane. For the
untwisted sectors we obtain:
\begin{eqnarray}
 |B,y=0,\eta\rangle &=& \frac{1}{(2\pi)^{5-r}}\prod_{i=6}^{9-s}\Big( \sum_{n_i} \mbox{
 e}^{i\frac{n_i}{R_i}\hat{q}}\Big) ~ \int d k^{5-r} ~ \mbox{
 e}^{ik\hat{q}} ~|B,k,\eta\rangle,
\end{eqnarray}
where with $R_l$ we indicate the value of the internal radii of the dimensions where the
brane extends, while with $R_i$ we refer to the transverse ones and, according to our
discussion, this localized bound state also contains a sum over the winding states.\\ For
the twisted sectors we get:
\begin{eqnarray}
 |B,y=0,\eta\rangle &=&   \frac{1}{(2\pi)^{5-r}}\int d k^{5-r} ~ \mbox{e}
 ^{ik\hat{q}} ~|B,k,\eta\rangle.
\end{eqnarray}

The invariance of the boundary state under the GSO and the orbifold projection always
requires that the physical boundary state is a linear combination of the two states
corresponding to $\eta=\pm$. In the conventions of \cite{9707068,9802088} these linear
combinations are of the form:
\begin{eqnarray}
 |B\rangle_{NS,U}   &=& \frac{1}{2}\Big( |B,+\rangle_{NS,U} ~-~|B,-\rangle_{NS,U}\Big)\\
 |B\rangle_{R,U}    &=& \frac{1}{2}\Big( |B,+\rangle_{R,U}  ~+~|B,-\rangle_{R,U}  \Big)\\
 |B\rangle_{NS,T}   &=& \frac{1}{2}\Big( |B,+\rangle_{NS,T} ~+~|B,-\rangle_{NS,T}\Big)\\
 |B\rangle_{R,T}    &=& \frac{1}{2}\Big( |B,+\rangle_{R,T}  ~+~|B,-\rangle_{R,T}  \Big)\\
\end{eqnarray}
A fractional D$p$ brane state can be written as:
\begin{eqnarray}
\label{ebound}
 |{\rm D}p\rangle &=& N_{U,p} ~
	\Big( |B\rangle_{NS,U} \pm |B\rangle_{R,U} \Big)
	\pm N_{T,p} ~
	\Big( |B\rangle_{NS,T} \pm |B\rangle_{R,T} \Big)
\end{eqnarray}

Now we want to determine the constants $N_{T,U}$ solving the open-closed consistency
condition. We have to compare the closed string cylinder diagram with the open string
one-loop diagram. We report the operator structure of this comparison in ((\ref{badil})$\ldots$(\ref{fine})),
while for the normalizations we obtain:
\begin{eqnarray}
 N_{U,p} &=& \frac{1}{2\sqrt{2}} ~ T_p ~ \sqrt{\frac{\prod_{l=1}^s (2\pi R_l)}{\prod_{i=1}^{n-s}(2\pi R_i)}}
~ \sqrt{\frac{1}{\phi^n}}\\
N_{T,p} &=& 2^{-s/2} ~ \sqrt{2\pi} ~ (4\pi^2 \ap)^{\frac{1-r}{2}}
\end{eqnarray}
where $$T_p=\sqrt{\pi}(4\pi^2 \ap)^{\frac{3-p}{2}}$$ and $\phi$ is the self dual volume.

\begin{eqnarray}
\label{badil} \langle NS\ NS\ +|\mbox{e}^{-lH_c}|NS\ NS\ +\rangle = \langle NS\ NS\
-|\mbox{e}^{-lH_c}|NS\ NS\ -\rangle\\ \no=\ \ (\frac{f_3}{f_1})^8\ \
=\ \ \mbox{Tr}_{NS}(\mbox{e}^{-2tH_o}),\\[2mm]
\no \langle NS\ NS\ +|\mbox{e}^{-lH_c}|NS\ NS\ -\rangle = \langle NS\ NS\
-|\mbox{e}^{-lH_c}|NS\ NS\ +\rangle\\
\no=\ \ (\frac{f_2}{f_1})^8\ \
=\ \ \mbox{Tr}_{R}(\mbox{e}^{-2tH_o}),\\[2mm]
\no \langle RR\ +|\mbox{e}^{-lH_c}|RR\ +\rangle = \langle RR\ -|\mbox{e}^{-lH_c}|RR\
-\rangle\\ \no =\ \ -(\frac{f_4}{f_1})^8\ \  =\ \ \mbox{Tr}_{NS}(\mbox{e}^{-2tH_o}(-)^F),\\[2mm]
\no \langle RR\ +|\mbox{e}^{-lH_c}|RR\ -\rangle = \langle RR\ -|\mbox{e}^{-lH_c}|RR\
+\rangle=\ \ 0,\\[2mm]
\label{NST} \langle NS\ NS;T\ +|\mbox{e}^{-lH_c}|NS\ NS;T\ +\rangle = \langle NS\ NS;T\
-|\mbox{e}^{-lH_c}|NS\ NS;T\ -\rangle\\ \no=\ \ (\frac{f_4f_3}{f_1f_2})^4\ \
=\ \ \mbox{Tr}_{NS}(\mbox{e}^{-2tH_o}\cdot g),\\[2mm]
\no \langle NS\ NS;T\ +|\mbox{e}^{-lH_c}|NS\ NS;T\ -\rangle = \langle NS\ NS;T\
-|\mbox{e}^{-lH_c}|NS\ NS;T\ +\rangle \\ \no=\ \ 0\ \
=\ \ -\mbox{Tr}_{R}(\mbox{e}^{-2tH_o}\cdot g),\\[2mm]
\label{RT}\langle RR;T\ +|\mbox{e}^{-lH_c}|RR;T\ +\rangle = \langle RR;T\
-|\mbox{e}^{-lH_c}|RR;T\ -\rangle\\ \no=\ \ -(\frac{f_3f_4}{f_1f_2})^4\ \
=\ \ \mbox{Tr}_{NS}(\mbox{e}^{-2tH_o}\cdot g\cdot (-)^F),\\[2mm]
\no \langle RR;T\ +|\mbox{e}^{-lH_c}|RR;T\ -\rangle = \langle RR;T\
 -|\mbox{e}^{-lH_c}|RR;T\ +\rangle \\ \label{fine}=\ \ 0\ \ = \ \ -\mbox{Tr}_R(\mbox{e}^{-2tH_o}\cdot g\cdot
(-)^F)
\end{eqnarray}
The definition of the function $f_i$ appearing here is as in \cite{anto}. Notice that the
identification of the twisted sectors is due to the relations (\ref{NST}) and (\ref{RT}).
Examining closely these relation, we see that this identification is very sensitive to the
sign of the $g$-orbifold projector on the ground state of the open-string and, moreover,
this sign can always be absorbed in the global normalization of the twisted sectors of the
boundary state. This ambiguity is characteristic only of the $\mathbb{Z}_2$ orbifold.

Now, we assume to make the opposite choice for the identification of the twisted sectors
(on the closed side of the open-closed string duality we exchange the RT sector with the
NST one). Then the tree level closed string amplitude
\begin{equation}
  \langle B|P|B\rangle
\end{equation}
continues to be zero, as it has to be. However it happens something particular at the
value (\ref{vol}) of the internal volume: the contributions from the twisted and untwisted
NS-NS sectors cancel between themselves. So, if we interpret the NS-NS-T sector as
contributing to the definition of an effective mass for the brane, as argued in the
paragraph (\ref{section}), we can say that at that particular value of the internal volume
(\ref{vol}) the fractional branes become massless, and this has been shown to be
consistent with the enhan\c{c}on mechanism.

The other choice of the sign, made frequently in the literature \cite{sen}, led instead to
suppose the existence of a non-BPS D$(r,s)$-truncated branes, (for $s=r=0$ you have the
famous non-BPS D-particle, \cite{particle}), that are stable for the value (\ref{vol}) of
the internal volume.

We report here, for generic $r$ and $s$, the calculation that determines the critical
volume:

\begin{eqnarray}
\no \langle B | P |B\rangle= \mbox{cost} ~ \Bigg[ \frac{\prod_{l=1}^s
(2\pi R_l)}{\prod_{i=1}^{4-s} (2\pi R_i)}\Big(\prod_{l=1}^{s}\sum_{n_l} \exp (-R^2 t\pi n_l^2/2\ap)\Big)\\
\no \Big(\prod_{l=1}^{n-s}\sum_{n_l} \exp (-\ap t\pi n_l^2/2R^2 )\Big) ~
\Big(\frac{f_3(q)^8-f_4(q)^8}{f_1(q)^8}\Big)- 2^4 (4\pi^2\ap)^{s-2} ~
\Big(\frac{f_2(q)^4f_3(q)^4}{f_1(q)^4f_4(q)^4}\Big)\Bigg]
\end{eqnarray}
where $q=\mbox{e}^{-\pi t}$. Using the relations
\begin{eqnarray}
\no\sum_l e^{-\pi t\ l^2}=f_1(e^{-\pi t}) f_3^2(e^{-\pi t})\\
\no f_4f_3f_2(e^{-\pi t})=\sqrt{2}
\end{eqnarray}
we get that this amplitude is zero for the following values of the radii:
\begin{equation}
R_l\ =\ \sqrt{\frac{\ap}{2}}\ \ \ \ \ \ \ \ \ \ R_i\ =\ \sqrt{2\ap}
\end{equation}
Note that this result is consistent with T-duality.\\
We see that for $s=0$ we get precisely the value (\ref{vol}) for the critical volume and
that it slightly changes for $s\neq 0$.

\newpage

\end{document}